\documentclass[twoside]{article}
\usepackage{qic,epsfig}
\usepackage{amsmath}

\newcommand{\ket}[1]{\vert #1\rangle}
\newcommand{\bra}[1]{\langle #1\vert}
\newcommand{\e}{\mathrm{e}}
\newcommand{\ii}{\mathrm{i}}

\begin{document}
\setlength{\textheight}{8.0truein}    

\runninghead{A QUANTUM REPEATER BASED ON DECOHERENCE FREE SUBSPACES}
            {U. DORNER, A. KLEIN, D. JAKSCH}

\normalsize\textlineskip
\thispagestyle{empty}
\setcounter{page}{1}


\vspace*{0.88truein}

\alphfootnote

\fpage{1}

\centerline{\bf {\bf A QUANTUM REPEATER BASED ON DECOHERENCE FREE SUBSPACES}}
\vspace*{0.37truein}
\centerline{\footnotesize UWE DORNER}
\vspace*{0.015truein}
\centerline{\footnotesize\it Clarendon Laboratory, University of Oxford, Parks Road}
\baselineskip=10pt
\centerline{\footnotesize\it Oxford OX1 3PU, United Kingdom}
\vspace*{10pt}
\centerline{\footnotesize ALEXANDER KLEIN}
\vspace*{0.015truein}
\centerline{\footnotesize\it Clarendon Laboratory, University of Oxford, Parks Road}
\baselineskip=10pt
\centerline{\footnotesize\it Oxford OX1 3PU, United Kingdom}
\vspace*{10pt}
\centerline{\footnotesize DIETER JAKSCH}
\vspace*{0.015truein}
\centerline{\footnotesize\it Clarendon Laboratory, University of Oxford, Parks Road}
\baselineskip=10pt
\centerline{\footnotesize\it Oxford OX1 3PU, United Kingdom}
\vspace*{0.225truein}


\vspace*{0.21truein}

\abstracts{
  We study a quantum repeater which is based on decoherence free
  quantum gates recently proposed by Klein {\it et al.} [Phys. Rev. A,
  {\bf 73}, 012332 (2006)]. A number of operations on the decoherence
  free subspace in this scheme makes use of an ancilla qubit, which
  undergoes dephasing and thus introduces decoherence to the system.
  We examine how this decoherence affects entanglement swapping and
  purification as well as the performance of a quantum repeater. We
  compare the decoherence free quantum repeater with a quantum
  repeater based on qubits that are subject to decoherence and show
  that it outperforms the latter when decoherence due to long waiting
  times of conventional qubits becomes significant. Thus, a quantum
  repeater based on decoherence free subspaces is a possibility to
  greatly improve quantum communication over long or even
  intercontinental distances.
}{}{}

\vspace*{10pt}

\vspace*{3pt}

\vspace*{1pt}\textlineskip    


\section{Introduction}
 Quantum communication is one of the experimentally most
advanced areas of quantum information processing and promises to
yield commercial applications in the near future~\cite{Gisin02}. In
addition to free space quantum communication, current setups mainly
use photon transmission in optical fibers and the distances over
which quantum cryptography is possible so far are in the range of up
to about 100km~\cite{Sergienko06}. However, quantum communication
over longer distances is primarily limited by photon loss, which
grows exponentially with the length of the fibre. A possible
solution of this problem is the use of quantum
repeaters~\cite{Briegel98,Duan01} to distribute maximally entangled
pairs of qubits over long distances. These pairs can then be used
for entanglement based quantum communication by
teleporting~\cite{Bennett93} quantum information from one party to
the other. The basic idea of a quantum repeater is to divide the
transmission line into shorter segments with a length of the order
of the attenuation length of the fibre. On each segment entangled
particle pairs are created and by applying entanglement
swapping~\cite{Zukowski93} and purification
protocols~\cite{Bennett96a,Deutsch96,Briegel98} entangled pairs of
larger distances are produced. Successive application of these steps
according to a nested repeater protocol~\cite{Briegel98} creates a
distant qubit pair with high entanglement fidelity.

During the purification process the entanglement fidelity of a pair
of qubits is successively increased by sacrificing auxiliary
entangled pairs. In the present paper we will use a purification
protocol known as ``entanglement pumping''~\cite{Duer99}. From a
practical point of view the use of entanglement pumping is
favourable compared to other purification schemes
~\cite{Bennett96a,Deutsch96,Bennett96b} since it requires
significantly fewer qubits and thus might be easier to implement.
However, the decrease of physical resources comes at the cost of
long operation times of the repeater during which quantum
information has to be stored. These waiting times grow quickly with
the distance of the two parties who desire to share an entangled
state and if they are too long the stored quantum information will
decohere to such a degree that the quantum repeater can not be
successfully operated anymore. This problem was recently addressed
by Hartmann {\it et al.}~\cite{Hartmann07} (see
also~\cite{Collins07}), who examined the limitations of a quantum
repeater (in terms of maximal distance) depending on the noise
strength. A number of modifications of the repeater protocol have
been proposed and it was shown that the maximal distance might be
increased by an order of magnitude at the cost of a reasonable
overhead of resources. Unlimited distances are only possible with
the help of quantum error correction. However, this imposes very
stringent error thresholds, which seem out of reach with present
technology.

In this paper we pursue the different and conceptually more
straightforward strategy of improving the quality of the quantum
memories at the repeater nodes. We study a repeater architecture
based on a scheme recently proposed by Klein {\it et
al.}~\cite{Klein06}, which relies on the concept of decoherence free
subspaces (DFSs)~\cite{Zanardi97,Duan98,Lidar98,Bacon00}. DFSs are a
method of passive error correction or error prevention and can
significantly increase the lifetime of quantum information and
reliability of quantum computing as already demonstrated in a number
of
experiments~\cite{Kielpinski01,Roos04,Roos06,Langer05,Fortunato02,Ollerenshaw03,Viola01,Kwiat00,Mohseni03,Bourennane04}.
In Ref.~\cite{Klein06} a logical qubit at a repeater node is
represented by two states of a decoherence free subspace of a
Hilbert space consisting of four atomic qubits. The logical qubits
are immune to collective noise thus greatly improving the lifetime
of stored quantum information. However, gate operations become more
complicated and slower than operations on ``bare'', unprotected
atomic qubits. In fact the two qubit operation proposed
in~\cite{Klein06} is not decoherence free since an unprotected
auxiliary qubit is used to mediate between two logical DFS qubits.
In the present work we examine how this noise affects the
performance of the quantum repeater network and show that it is
possible to create entangled pairs of high fidelity over
intercontinental distances. We compare the results to a repeater
based on unprotected qubits, i.e., qubits that are subject to
decoherence, and show that it outperforms the latter when
decoherence due to long waiting times becomes significant. Although
we consider only the special case of a quantum repeater, we
emphasise that the methods we employ are generally applicable to
arbitrary quantum networks.

This paper is organised as follows. In Sec.~\ref{Sec:IIrep} we
briefly review the quantum repeater protocol. In
Sec.~\ref{Sec:IIIrep} we present the error model we use for quantum
repeaters based on unprotected qubits and DFS qubits. In the same
section we furthermore describe the quantum circuits necessary to
implement the repeater. In Sec.~\ref{Sec:IVrep} we present the
results of simulations for both repeater setups. Finally we conclude
in Sec.~\ref{Sec:Conclrep}.

\section{The quantum repeater}
\label{Sec:IIrep}  In this paper we use the nested repeater protocol
developed in~\cite{Briegel98,Duer99}, which consists of a
combination of entanglement purification and entanglement swapping.
The goal is to create a highly entangled pair between two parties,
say, $A$ and $B$, which might be attempted by transmitting a photon
through a fibre. However, unwanted noise will decrease the
entanglement fidelity of the qubit pair monotonously with the
distance between $A$ and $B$. The fidelity can be increased again by
applying purification procedures, in which additional entangled
pairs between $A$ and $B$ are created and sacrificed in order to
distill an entangled pair with high fidelity. However, if the
distance between $A$ and $B$ is too large, the entanglement fidelity
of the pairs can drop below a minimum value $f_{\mathrm{min}}$,
which is required by the purification protocol to increase the
entanglement fidelity~\cite{Duer99}.  To overcome this problem, a
number of intermediate nodes $N_i$ with sufficiently small distances
$l_0$ are introduced between $A$ and $B$, and entangled qubit pairs
are prepared between each of the intermediate nodes such that the
entanglement fidelity of each pair is greater than
$f_{\mathrm{min}}$. These pairs can be purified and connected via
entanglement swapping to create an entangled pair of larger
distance. For the setup used in this work we consider purification
via ``entanglement pumping''~\cite{Duer99}, which requires
considerably less qubits than other
schemes~\cite{Bennett96a,Deutsch96,Bennett96b} and is thus
preferable from a practical point of view.
\begin{figure}[t!]
\centerline{\epsfig{file=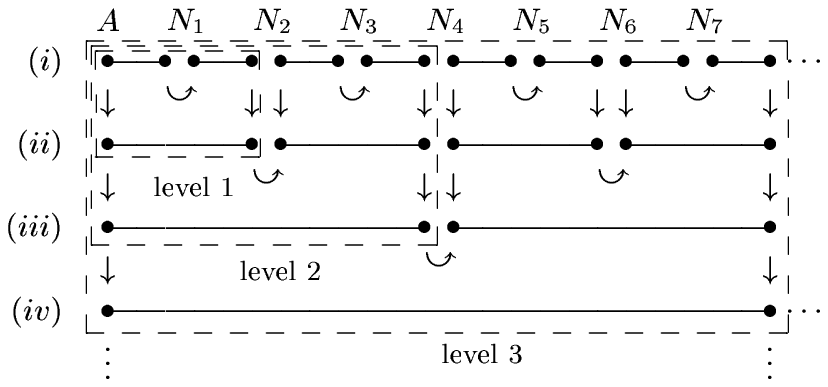}} \vspace*{13pt}
\fcaption{%
Illustration of the nested repeater protocol. For explanation see
text.}\label{fig:schemerep}
\end{figure}

The principle of the nested repeater protocol is illustrated by the
example shown in Fig.~\ref{fig:schemerep}: Each bullet represents a
qubit and the lines between them indicate entanglement. On repeater
level 1 (indicated by the dashed box on the top left) two entangled
qubit pairs between nodes $A$ and $N_1$ and between nodes $N_1$ and
$N_2$ are created in line $(i)$. The two qubits at node $N_1$ are
then connected via entanglement swapping, creating an entangled pair
of qubits with larger distance, indicated by the curved arrow.
Within the schematic of Fig.~\ref{fig:schemerep} the state of this
qubit pair is then transfered by a quantum operation to the qubit
pair in line $(ii)$. We then generate further entangled pairs in
line $(i)$, which are used to purify the pair in line $(ii)$,
indicated by vertical arrows. Given that $l_0$ is the distance
between the nodes, the final pair will have a distance $2l_0$.
Repeater level 2 consists of two adjacent level 1 repeaters, both of
which create an entangled pair of distance $2l_0$ on line $(ii)$.
These two pairs are then connected, the resulting pair is transfered
to line $(iii)$ and subsequently produced pairs are used to purify
the pair in line $(iii)$, which has now a distance $4l_0$. Repeater
level 3 is constructed in the same way: We use two level 2 repeaters
which successively generate pairs in line $(iii)$. After connecting
them the resulting pair is transfered to line $(iv)$, and
subsequently produced pairs are used to purify the pair with
distance $8l_0$ in line $(iv)$. In this example the distance between
the entangled qubits is doubled with each further repeater level. In
general, the number of entanglement swapping steps on each repeater
level can vary from level to level and is adapted to the specific
physical situation. The distance between entangled qubits on
repeater level $n$ is then given by
\begin{equation}
S_n \equiv l_0\prod_{j=1}^n (L_j+1)
\end{equation}
for $n\ge1$ and $S_0 = l_0$. The quantity $L_j$ is the number of
connections on level $j$ immediately before the final qubit pair of
this level is purified, i.e.~in the example shown in
Fig.~\ref{fig:schemerep} we have $L_1=L_2=L_3=1$. Thus, $L_j$ is
generally different from the total number of connections necessary
to operate a level $j$ repeater.

We note that in basically all quantum communication schemes flying
qubits are represented by photons. For our setup we assume that the
states of these photons are first transferred to stationary qubits
(atoms) creating an entangled pair with entanglement fidelity $f_0$,
which might be lower than the entanglement fidelity of the photon
pair. We use this fidelity $f_0$ as the starting fidelity in all our
discussions. We also note that the state transfer described above is
not necessary if we do an appropriate relabeling of the qubits.
However, for reasons given in Sec.~\ref{Sec:IVrep} we assume that
this transfer is done via a quantum operation.

\section{Error models}
\label{Sec:IIIrep}

\subsection{Error model for unprotected qubits}\label{Sec:IIIArep}

The error model we use in this paper is motivated by realisations of
quantum information processing with single atoms stored in tight
traps \cite{Roos04,Roos06,Klein06}. In this scenario, qubits can be
represented by two metastable states of atoms. The lifetime of these
states is typically on the order of several minutes or longer, such
that their spontaneous decay can be neglected
\cite{Kielpinski01,Roos04,Roos06,Langer05,Kuhr05}. The major source
of decoherence is then given by dephasing, represented by the
$\sigma_z$-Pauli operator. Unitary single qubit operations necessary
to manipulate the qubits can be realised by laser pulses and static
magnetic or electric fields. It is thus easily possible to implement
Hamiltonians which are proportional to Pauli operators such that the
time evolution of the system corresponds to rotations around the
$x,y,$ and $z$ axis of the Bloch sphere. For two qubit operations
various schemes have been developed~\cite{Cirac04}. Here, we
consider two qubit operations caused by an Ising interaction, which
can be realised via the collisional interaction between neutral
atoms stored in optical traps~\cite{Jaksch-PRL-99,Mandel03}.

\subsubsection{Single qubit gates and measurements}\label{Sec:IIIA1rep}

We assume that the major source of noise is dephasing so that the
time evolution of the system state $\rho$ whilst applying a gate
described by $H_\alpha^i$ on qubit $i$ is determined by the master
equation ($\hbar=1$)
\begin{equation}
\dot\rho = -\ii[H_\alpha^i,\rho] +
\frac{\gamma}{2}(\sigma_z^i\rho\sigma_z^i - \rho) \,.
\label{eq:master1rep}
\end{equation}
Here,
\begin{equation}
H_\alpha^i = \Omega_\alpha\sigma_\alpha^i  ,
\end{equation}
where $\sigma_\alpha^i$ are the Pauli operators with $\alpha =
0,x,y,z$. In the case of $\alpha=0$ we set $\sigma^i_0 = \mathbf{1}$
so that the above master equation also describes the dephasing of a
quantum channel or memory. For simplicity we furthermore assume that
$\Omega_\alpha$ is real and non-negative. For vanishing noise (i.e.,
$\gamma=0$) the $i$th qubit undergoes a rotation around the $x$, $y$
or $z$ axis
\begin{equation}
\rho\rightarrow R_\alpha^i(\theta)\rho R_\alpha^i(\theta)^\dagger
\quad\text{with}\quad  R_\alpha^i(\theta) \equiv
\e^{-\ii\frac{\theta}{2}\sigma_\alpha^i},
\end{equation}
where the rotation angle is given by $\theta=2\Omega_\alpha t$.

In the presence of dephasing (i.e., $\gamma\ne 0$) the solutions of
Eq.~(\ref{eq:master1rep}) can be described by quantum operations
$\mathcal{E}_\alpha^i$,
\begin{equation}
\rho\rightarrow \mathcal{E}_\alpha^i(\theta)[\rho] = \sum_k
E_{\alpha,k}^i  \rho \left(E_{\alpha,k}^i\right)^\dagger.
\end{equation}
For $\alpha=0$, i.e., if the commutator in Eq.~(\ref{eq:master1rep})
vanishes, the evolution of the density operator is given by
\begin{equation}
\rho\rightarrow \mathcal{E}_0^i(\gamma t)[\rho] =
 p_1(\gamma t)\rho + p_2(\gamma t)\sigma_z^i\rho\sigma_z^i ,
\end{equation}
where
\begin{equation}
p_1(\gamma t) = \frac{1}{2}\left( 1+\e^{-\gamma t} \right),\quad
p_2(\gamma t) = \frac{1}{2}\left( 1-\e^{-\gamma t} \right).
\label{eq:p_gammarep}
\end{equation}
Note that for $\alpha=z$ the noise operator $\sigma^i_z$ commutes
with $H_z^i$ so that
\begin{equation}
\mathcal{E}_z^i(\theta)[\rho] = \mathcal{E}_0^i(\gamma
t)[R_z(\theta)\rho R_z(\theta)^\dagger]
 = R_z(\theta)\mathcal{E}_0^i(\gamma t)[\rho]R_z(\theta)^\dagger,
\end{equation}
i.e., the process can be replaced by a perfect rotation followed by
noise in the channel, or vice versa. In the remainder of the present
work we omit the superscript $i$ whenever it is clear from the
context (e.g.~in quantum circuits) on which qubit the operation is
acting on.

Non-ideal measurements are described in this paper by the positive
operator valued measure~\cite{Duer99}
\begin{eqnarray}
P_0 &=& \eta\ket{0}\bra{0} + (1-\eta)\ket{1}\bra{1} \\
P_1 &=& \eta\ket{1}\bra{1} + (1-\eta)\ket{0}\bra{0}
\end{eqnarray}
with $0\le \eta \le 1$. The parameter $\eta$ is the probability to
obtain the correct result if a measurement is done in the
$\{\ket{0},\ket{1}\}$-basis.

\subsubsection{Two qubit gates}\label{Sec:IIIA2rep}

We consider two qubit operations mediated by an Ising interaction.
The dynamics of the system can thus be described by a master
equation of the form
\begin{equation}
\dot\rho = -\ii[\Omega_{zz}\sigma_z^i\sigma_z^j,\rho] +
\frac{\gamma}{2}(\sigma_z^i\rho\sigma_z^i - \rho) +
\frac{\gamma}{2}(\sigma_z^j\rho\sigma_z^j - \rho) \,,
\label{eq:isingrep}
\end{equation}
where we assume that the noise on qubit $i$ and $j$ is uncorrelated.
Since the Ising Hamiltonian commutes with the noise operators
$\sigma_z^i$, the time evolution is simply given by
\begin{align}
  \rho\rightarrow \mathcal{E}_{zz}^{ij}(\xi)[\rho] = &
  \mathcal{E}_{0}^i(\gamma t)[ \mathcal{E}_{0}^j(\gamma t)[
  \e^{-\ii\frac{\xi}{2}\sigma_z^i\sigma_z^j}\rho
  \e^{\ii\frac{\xi}{2}\sigma_z^i\sigma_z^j} ] ]  \,,
\end{align}
where $\xi = 2\Omega_{zz}t$. The order of the three distinct
operations in the above equation is arbitrary.  This means that
$\mathcal{E}_{zz}^{ij}$ can, for example, be described by a perfect
operation followed by dephasing in the quantum channels.
\begin{figure}[ht]
\centering\includegraphics[]{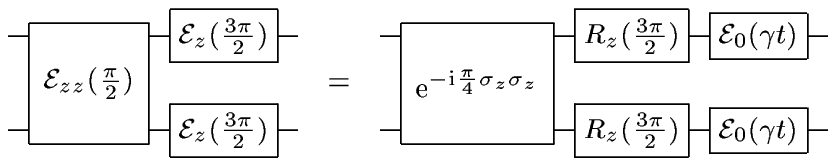} \vspace*{13pt}
\fcaption{%
  Controlled-$Z$ operation, $\mathcal{Z}^{AA}_\downarrow$, on two unprotected
  qubits. The left circuit can be replaced by an effective circuit
  consisting of ideal gates followed by noise in the quantum channels.
  The time $t$ is given by $t = \pi/4\Omega_{zz} + 3\pi/4\Omega_z$,
  which is the time needed to perform the noiseless gates.}
\label{fig:CZrep}
\end{figure}
An application of this fact is illustrated in Fig.~\ref{fig:CZrep},
which shows the realisation of a noisy controlled-$Z$ gate on two
qubits. In the following figures we denote this gate as
$\mathcal{Z}_\downarrow^{AA}$, the superscript indicating that the
gate is acting on two unprotected (atomic) qubits. The subscript
defines control and target qubit, i.e., in quantum circuits the
arrowhead points to the target qubit. Strictly speaking this is not
necessary since this gate is symmetric under qubit exchange.
However, in later sections we use a similar notation for
controlled-$(-Z)$ gates which are not symmetric.

\subsubsection{Building blocks of the quantum repeater using unprotected qubits}\label{Sec:IIIA4rep}

\begin{figure}[ht]
\centering\includegraphics[]{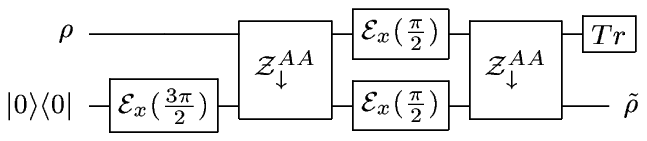} \vspace*{13pt}
\fcaption{%
  State transfer between two unprotected qubits. In the noiseless case the output state
  $\tilde\rho$ would be equal to the input state $\rho$.}
\label{fig:transfer_arep}
\end{figure}

\begin{figure}[t]
\centering\includegraphics[]{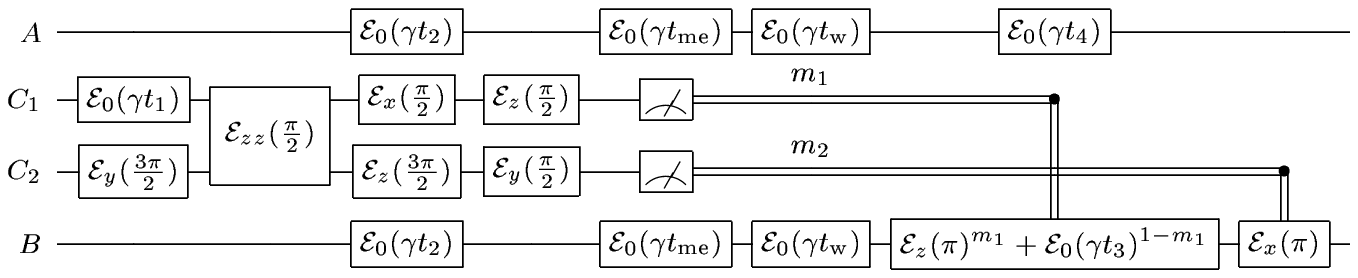} \vspace*{13pt}
\fcaption{%
  Entanglement swapping of qubit pairs $A-C_1$ and $C_2-B$. The state
  of qubit $C_1$ is teleported to qubit $B$ by using the entanglement
  of qubit pair $C_2-B$ yielding an entangled pair $A-B$. The waiting
  times are given by $t_1 = 3\pi/4\Omega$, $t_2 = 5\pi/4\Omega +
  \pi/4\Omega_{zz}$ and $t_3 = \pi/2\Omega$.  The value of $t_4$
  depends on the outcome of the measurement of qubit $C_2$, given by
  $m_2=0,1$. In particular we have $t_4 = \pi/\Omega$ if $m_2=1$ and
  $t_4 = \pi/2\Omega$ if $m_2=0$. The time $t_\mathrm{me}$ corresponds to the
  time necessary to perform a measurement and $t_\mathrm{w} = S_{n-1}/c$ is the
  classical communication time between qubits $C_{1,2}$ and qubit $B$
  where $S_{n-1}$ is the distance between $C_{1,2}$ and $B$ on
  repeater level $n$.}
\label{fig:teleportation_arep}
\end{figure}

\begin{figure}[t]
\centering\includegraphics[]{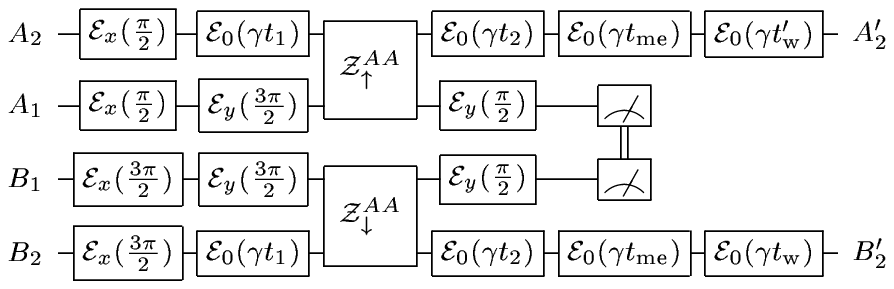} \vspace*{13pt}
\fcaption{%
  Quantum circuit for an entanglement purification step. The waiting
  times are given by $t_1 = 3t_2 = 3\pi/4\Omega$, $t_\mathrm{w} = S_n/c$, where
  $S_n$ is the distance between the qubits $A_{1,2}$ and $B_{1,2}$ on
  repeater level $n$, $t_\mathrm{me}$ is the time required to perform a
  measurement and $t'_\mathrm{w} = t_\mathrm{w} + \pi/2\Omega$.}
\label{fig:purif_arep}
\end{figure}

As indicated in Sec.~\ref{Sec:IIrep}, three basic modules are needed
to run the quantum repeater. In particular, these are the transfer
of a state from one qubit to another one, entanglement swapping, and
entanglement purification.  In
Figs.~\ref{fig:transfer_arep}-\ref{fig:purif_arep} we show possible
implementations of these three blocks according to the error models
and gate operations described in Secs.~\ref{Sec:IIIA1rep}
and~\ref{Sec:IIIA2rep}. In our simulations qubits that are measured
are immediately removed from the system by tracing them out. The
removal of a qubit is indicated by a $Tr$-symbol in the
corresponding quantum circuits unless we perform measurements that
classically control further operations. Furthermore, we assume for
simplicity that $\Omega \equiv \Omega_x = \Omega_y = \Omega_z$.

Clearly, the partitioning of the three blocks into elementary gate
operations is not unique and in the presence of noise different
architectures can lead to different fidelities. In our simulations
we compared various possibilities, and the realisations shown in
this article are the ones which led to the best results for the
error models and corresponding error parameters we use, see
Sec.~\ref{Sec:IVrep}. Moreover, in the case of a quantum repeater
based on unprotected qubits, which is used for communication over
long distances, the dominant source of noise is due to long waiting
times during classical communications (see below and
Ref.~\cite{Hartmann07}) and the specific partitioning of the blocks
becomes less important.

Apart from noise during gate operations, the quantum circuits shown
in this section also include noise which is due to waiting times of
qubits. Whenever it is unavoidable that a qubit has to wait until an
operation on another qubit is finished or until a classical signal
arrives it undergoes dephasing $\mathcal{E}_0$.
Fig.~\ref{fig:teleportation_arep} shows entanglement swapping
between two entangled pairs $A-C_1$ and $C_2-B$ of qubits. The goal
is to teleport the state of qubit $C_1$ to qubit $B$ by using the
entanglement of the pair $C_2-B$. In a noiseless version of this
circuit qubit $A$ would be simply represented by a straight line,
which is disconnected from the remaining qubits, and undergoes no
operations (and thus it would normally be omitted in the circuit).
However, in the presence of noise, it has to wait until the whole
procedure is finished and undergoes dephasing during this time. On
repeater level $n$ we have to wait a time $t_\mathrm{w} =
S_{n-1}/c$, where $c$ is the speed of light, until the classical
signal resulting from the measurement of qubits $C_1$ and $C_2$
arrives at qubit $B$. For large distances between qubits $C_1,\,C_2$
and $B$, i.e. on higher repeater levels, this waiting times will be
quite long. For instance, taking $S_{n-1}=1000$km yields a waiting
time $t_\mathrm{w}\approx 3$ms, which is considerably larger than
gate operation times of atomic qubits that are typically in the
$\mu$s regime, see for instance \cite{Mandel03,Riebe06}.
Realisations solely based on solid state systems such as electron
spins in quantum dots would have even shorter gate operation
times~\cite{Burkard04}. However, in the case of a quantum repeater
this is of no advantage since in solid state systems coherence times
are typically shorter than in atomic systems where it can exceed
$100$ms~\cite{Kuhr03,Kuhr05}. Since the waiting times during
classical communication necessary for entanglement swapping are
independent of the implementation, solid state realisations would be
less suitable.

If more than two qubit pairs are to be connected we can use a
simultaneous entanglement swapping scheme. For example three entangled
qubit pairs $A-C_1$, $C_2-C_3$ and $C_4-B$, can be transformed into
one entangled qubit pair $A-B$ by teleporting the state of qubit $C_2$
to qubit $A$ (using the entanglement of $A-C_1$) and by teleporting the
state of qubit $C_3$ to qubit $B$ (using the entanglement of $C_4-B$)
at the same time. In general, a sequence of qubit pairs $A-C_1,\,
C_2-C_3,\,\ldots,\, C_{2L_n}-B$ with $L_n$ even can be transformed into
a single entangled qubit pair $A-B$ by simultaneously teleporting the
state of qubit $C_{L_n}$ to $C_{L_n-2}$ and the state of qubit
$C_{L_n+1}$ to $C_{L_n+3}$ before $C_{L_n-2}$ is teleported to
$C_{L_n-4}$ and $C_{L_n+3}$ to $C_{L_n+5}$ and so on. If $L_n$ is odd
we start by teleporting $C_{L_n-1}$ to $C_{L_n-3}$ and $C_{L_n}$ to
$C_{L_n+2}$ before $C_{L_n-3}$ is teleported to $C_{L_n-5}$ and
$C_{L_n+2}$ to $C_{L_n+4}$ and so on. This leads to two remaining
entangled qubit pairs $A-C_{2L_n-1}$ and $C_{2L_n}-B$ which can be
connected to a single pair $A-B$ via the procedure shown in
Fig.~\ref{fig:teleportation_arep}. In total this method takes a time
$\lceil L_n/2 \rceil(t_{\mathrm{sw}} + S_{n-1}/c)$ where $t_{\mathrm{sw}}$ is
the time of the operation shown in Fig.~\ref{fig:teleportation_arep}
minus the classical communication time $t_\mathrm{w}$ and $\lceil.\rceil$
is the ceiling function.

Fig.~\ref{fig:purif_arep} shows the quantum circuit for an
entanglement purification step of a qubit pair. The circuit
corresponds to the scheme proposed by Deutsch {\it et
al.}~\cite{Deutsch96}, but here it is expressed in terms of
operations which correspond to our gate and error model. We start
with two entangled pairs $A_1-B_1$ and $A_2-B_2$ and sacrifice the
pair $A_1-B_1$ in order to get---whenever we obtain a coincidence in
the measurements---a new pair $A'_2-B'_2$, which can have a higher
entanglement fidelity than the pair $A_2-B_2$. Whether the fidelity
increases depends on the noise strength and the fidelity of the
input pairs and will be discussed in Sec.~\ref{Sec:IVrep}. If we do
not get coinciding measurement results the procedure fails and has
to be repeated with a new set of pairs. The measurement result has
to be classically exchanged between node $A$ (the location of qubits
$A_{1,2}$) and node $B$ (the location of qubits $B_{1,2}$), which
are a macroscopic distance apart from each other. This is indicated
by the double wire connecting the two measurements in
Fig.~\ref{fig:purif_arep}. The classical communication time is given
by $t_\mathrm{w} = S_n/c$ and will thus be, as in the case of
entanglement swapping, quite large on higher repeater levels leading
to a significant dephasing of qubit $A'_2$ and $B'_2$.

In addition to the already discussed waiting times during
entanglement swapping and purification there will be further waiting
times for entangled qubits during the repeater protocol: After an
entangled pair on repeater level $n$ is created we have to wait a
certain time until a second pair is available that we can use for a
purification step. In the following, we derive a lower bound for
these waiting times.

On the lowest repeater level photons are sent to the repeater nodes
and their state is transfered to atomic qubits. The traveling time of
the photons can be omitted since they can be triggered such that they
arrive at the nodes just in time before a transfer is possible.  After
the transfer, which consumes a time $t_0$, entanglement swapping is
performed $L_1$ times taking a total time $t_0^\prime = t_0 + \lceil
L_1/2\rceil (t_{\mathrm{sw}} + S_0/c)$, where we take
$t_{\mathrm{sw}}=9\pi/4\Omega+\pi/4\Omega_{zz} + t_{\mathrm{me}}$. The
resulting state is then transfered to another qubit pair in a time
$t_{\mathrm{tr}}=5\pi/2\Omega+\pi/2\Omega_{zz}$. Only after this
transfer is complete, it is possible to transfer photonic states to
atomic qubits again and connect them. These pairs are then used to
purify the pair previously generated, which takes a time
$t_{\mathrm{pur}}=5\pi/2\Omega+\pi/4\Omega_{zz} + t_{\mathrm{me}}$ for
the operations and measurements, and a time $S_1/c$ for the classical
communication of the measurement outcome. During this communication,
we can already start creating a subsequent pair necessary for
purification, such that the second purification step can be started
after a delay $t_{\mathrm{pur}}+\max(t_0^\prime, S_1/c)$. The
purification is performed $K_1$ times and thus the minimum time it
takes to create a pair on repeater level 1 is given by
\begin{equation}
  t_1 = t_{\mathrm{tr}} + t_0^\prime + K_1
  \left[t_{\mathrm{pur}} + \max \left( t_0^\prime\,,\,S_1/c
   \right) \right] \,.
\end{equation}
During this process, the pairs on level 1 might have to wait until
subsequent pairs for purification are created. After each
purification step, this additional waiting time is given by
$t_\mathrm{aw} = \max(0,t_0^\prime - S_1/c)$.

On higher levels, the minimum additional waiting time can be
estimated as follows. In order to create a pair on level $n \geq 2$
one has to create $\prod_{m=2}^n  (K_m+1)$ times the pairs on level
1, which takes at least a time
\begin{equation}
t_n = t_1 \prod_{m=2}^n  (K_m+1). \label{eq:optimerep}
\end{equation}
However, while the pairs created on level $n-1$ are used to purify
the pairs on level $n$, we can already start to prepare the pairs on
level $n-2$, $n-3$, an so on. Hence, the minimum time the pairs on
level $n$ have to wait for the completion of the pairs on level
$n-1$ is given by
\begin{equation}
  t^{\mathrm{c}}_{n-1} = \max\left\{0\,,\, t_{n-1}
 - \sum_{l=1}^{n-2} t_l \right\}
\end{equation}
for $n\ge2$ and $t_0^c=t_0$. The minimum additional waiting time
after each purification and transfer step on level $n$ is then given
by
\begin{equation} \label{Eq:Waitpurrep}
  t_\mathrm{aw} = \max\left\{0\,,\, \lceil L_n/2\rceil (t_\mathrm{sw} + S_{n-1}/c) +
  t^{\mathrm{c}}_{n-1} -  S_n/c   \right\} \,
\end{equation}
and
\begin{equation}\label{Eq:Waittrrep}
  {\tilde t}_\mathrm{aw} = \lceil L_n/2\rceil (t_\mathrm{sw} + S_{n-1}/c) +
  t^{\mathrm{c}}_{n-1}  \,,
\end{equation}
respectively. Note that in Eq.~(\ref{Eq:Waitpurrep}) the waiting
time $t_\mathrm{w} = S_{n}/c$ has been subtracted from the
additional waiting time $t_\mathrm{aw}$, since it is already
included in the purification scheme, see Fig.~\ref{fig:purif_arep}.

Equation (\ref{eq:optimerep}) can be used to estimate the operation
time for the quantum repeater. It provides only a lower bound since
it gives the time if the operation of the repeater was successful
``in one go'', i.e., if all involved purification steps have been
successful. However, the probability for this to happen is extremely
small~\cite{Hartmann07} and most likely one would operate the
repeater in a different way, such that on each repeater level one
would wait until the corresponding purification steps are
successful, which introduces further waiting times.

\subsection{Error model for DFS qubits}\label{Sec:IIIBrep}


The setup we consider for an experimental implementation of the
repeater nodes utilises single atoms stored in neighbouring dipole
traps, such as the wells of an optical lattice \cite{Klein06}.  Recent
experiments \cite{Kielpinski01,Roos04,Roos06,Langer05,Kuhr05} showed
that for this case the coupling of the qubits to their environment,
for example caused by electric or magnetic stray fields, can be
considered to be homogeneous, that means identical for all qubits.
Thus it is possible to extend the lifetime of the stored information
considerably by encoding it in a DFS, which protects the qubits from
homogeneous noise. In the next subsection, we briefly discuss two
possible DFSs which we consider in this paper. The first DFS scheme,
which was used in Ref.~\cite{Klein06}, encodes a logical qubit in four
two-level atoms in such a way that it is protected against arbitrary
kinds of homogeneous noise \cite{Bacon00}. The second DFS scheme
employs only two atoms, and is therefore easier to realise, but
protects the encoded quantum information only against homogeneous
dephasing. However, this is sufficient for a lot of implementations,
as has already been demonstrated in experiments
\cite{Kielpinski01,Roos04,Roos06,Langer05}. For both cases we present
how the single and two qubit gates necessary for implementing the
repeater protocol can be performed and which limitations occur.

\subsubsection{Single qubit gates, two qubit gates and measurements of DFS qubits}
\label{SecIIIB1rep}
The two logical states of a DFS qubit are represented by two states
of a system consisting of four two-level atoms, which are stored in
an array of dipole traps,
\begin{equation}
\begin{split}
 \ket{0}_{\mathrm{DFS}} &= \frac{1}{2}(\ket{01} - \ket{10})
 \otimes (\ket{01} - \ket{10}) \,,\\
 \ket{1}_{\mathrm{DFS}} &= \frac{1}{2\sqrt3}
 (2\ket{1100} + 2\ket{0011} - (\ket{01} + \ket{10})^{\otimes2}) \,,
\label{eq:logical_qubitrep}
\end{split}
\end{equation}
where $\ket{ijkl}=\ket{i}_1\ket{j}_2\ket{k}_3\ket{l}_4$ with
$i,j,k,l=0,1$ are the basis states of the four-atom system. This
subspace does not couple to collective noise corresponding to
fluctuating fields of the form
\begin{equation}
H_{I} = \sum_{i=1}^4 \sigma^x_i B_x + \sigma^y_i B_y + \sigma^z_i
B_z \,. \label{decoherence_fieldrep}
\end{equation}
As a consequence, the subspace is immune to all kinds of homogeneous
noise. So far the DFS in equation (\ref{eq:logical_qubitrep}) has
not been implemented using an atomic system. There are, however,
experimental realisations of a DFS which protects qubits against
homogeneous dephasing \cite{Kielpinski01,Roos04,Roos06,Langer05} using the simpler DFS described below. It
has been demonstrated that the coherence time of quantum information
stored in such a DFS is ultimately limited by the lifetime of the
excited atomic level with respect to spontaneous
decay~\cite{Roos04,Roos06}. The lifetime of ground state hyperfine
levels with respect to spontaneous decay is extremely long, in fact
times exceeding 10 minutes have been observed~\cite{Langer05}. We
can therefore neglect uncorrelated spontaneous emission and assume
that quantum information is stored without loss in the DFS.
\begin{figure}[t]
\centering\includegraphics[]{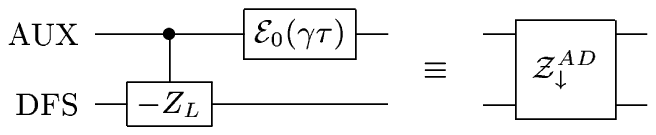} \vspace*{13pt}
\fcaption{%
  Noisy controlled-($-Z$) operation, $\mathcal{Z}^{AD}_\downarrow$,
  between an auxiliary qubit and a DFS qubit. In reality the
  dephasing takes place during the gate operation. In simulations it
  can be applied before or after the controlled-($-Z$) operation.}
\label{fig:CZAUXrep}
\end{figure}

It was shown in~\cite{Klein06} that single qubit rotations
$R_x(\theta)$ and $R_z(\theta)$ can be done without leaving the
decoherence free subspace and thus we assume that these operations are
performed without any error. In contrast to this it was shown that a
feasible implementation of a two qubit gate can be achieved by
applying a controlled-($-Z$) operation, which involves the use of an
unprotected auxiliary atom. The five atoms are then subject to
collective noise of the form $H_I + \sigma^x_{\mathrm{AUX}} B_x +
\sigma^y_{\mathrm{AUX}} B_y + \sigma^z_{\mathrm{AUX}} B_z$ and the
action of this operator on a state $\ket{\psi}_\mathrm{DFS}
\ket{\phi}_\mathrm{AUX}$ is given by $\ket{\psi}_\mathrm{DFS}
\ket{\phi}_\mathrm{AUX} \rightarrow \ket{\psi}_\mathrm{DFS}
(\sigma^x_{\mathrm{AUX}} B_x + \sigma^y_{\mathrm{AUX}} B_y +
\sigma^z_{\mathrm{AUX}} B_z) \ket{\phi}_\mathrm{AUX}$.  Hence, we can
assume that the noise acts independently on the auxiliary atom.  We
further restrict our considerations to dephasing noise
$B_z\sigma^z_\mathrm{AUX}$. The dynamics of the auxiliary atom is thus
described by the model detailed in Sec.~\ref{Sec:IIIrep}. Since the
noise operation (dephasing of the auxiliary atom) and the
controlled-($-Z$) operation commute, the combined operation can be
represented by an effective operation as shown in
Fig.~\ref{fig:CZAUXrep}, where $\tau$ is the time needed to perform
the controlled-($-Z$) gate.
In the following figures we denote this gate as
$\mathcal{Z}_\downarrow^{AD}$, the arrow again defining control and
target qubit. Since $\tau$ is relatively large ($\sim 1\,$ms) we
expect decoherence caused by the auxiliary atom to be the major
limiting factor in our setup, because the long waiting times of the
qubits during classical communication between the repeater nodes do
not play any role for the DFS qubits.

Although Ref.~\cite{Klein06} concentrates on a DFS given by
Eq.~(\ref{eq:logical_qubitrep}), we point out that in the case where
only collective dephasing (i.e. $B_x = B_y = 0$) is the relevant
source of noise a DFS consisting of two atoms is sufficient. In this
case, the logical states are
\begin{equation}
\begin{split} \ket{0}_\mathrm{DFS} = \frac{1}{\sqrt2}(\ket{01} +
\ket{10}) \,, \\
 \ket{1}_\mathrm{DFS} = \frac{1}{\sqrt2}(\ket{01} -
\ket{10}) \,. \label{eq:logical_qubit2rep}
\end{split}
\end{equation}
The controlled-($-Z$) operation between an auxiliary (atomic) qubit
and a DFS qubit as well as rotations $R_z(\theta)$ can be done in
exactly the same way as described in Ref.~\cite{Klein06} (omitting
two of the four atoms) with the same fidelities and operation times.
Rotations $R_x(\theta)$ can be performed using a laser to induce a
rotation around the $z-$axis of the Bloch sphere of, e.g., the first
atom constituting the DFS. As in the case of the four-qubit DFS,
these operations can be done without leaving the DFS and noise is
mainly introduced by the auxiliary atom.  Hence, all methods and
results presented in this paper are also valid for the DFS spanned
by the states given in Eq.~(\ref{eq:logical_qubit2rep}).
\begin{figure}[t]
\centering\includegraphics[]{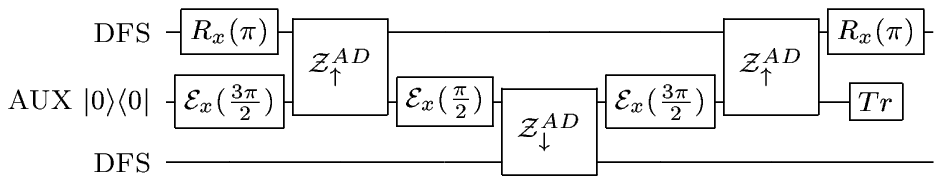} \vspace*{13pt}
 \fcaption{%
  Implementation of a controlled-($-Z$) operation,
  $\mathcal{Z}^{DD}_\downarrow$, between two DFS qubits.}
\label{fig:CMPHASErep}
\end{figure}
\begin{figure}[t]
\centering\includegraphics[]{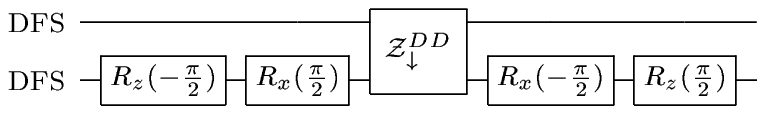} \vspace*{13pt}
\fcaption{%
  Quantum circuit for a controlled-not operation,
  $\mathcal{X}^{DD}_\downarrow$, between two logical qubits.}
\label{fig:CNOTrep}
\end{figure}
\begin{figure}[t]
\centering\includegraphics[]{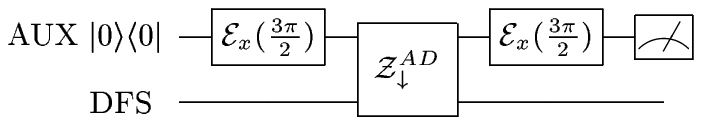}
\fcaption{%
  Measurement of a DFS qubit.}
\label{fig:measurerep}
\end{figure}

The operation depicted in Fig.~\ref{fig:CZAUXrep} can be used to
implement controlled operations between two DFS qubits by using an
auxiliary qubit. Fig.~\ref{fig:CMPHASErep} shows a possibility to
implement a controlled-($-Z$) gate between two DFS qubits. The
single qubit operations and the measurement of the auxiliary qubit
are the same as in Sec.~\ref{Sec:IIIArep}.  The controlled-($-Z$)
operation can then be used to generate a controlled-not between two
DFS qubits as shown in Fig.~\ref{fig:CNOTrep}. Analogously to our
previous notation, these are denoted in the following figures as
$\mathcal{Z}^{DD}_\downarrow$ and $\mathcal{X}^{DD}_\downarrow$.

In order to measure the state of a DFS qubit we again make use of an
auxiliary qubit. The corresponding circuit is shown in
Fig.~\ref{fig:measurerep}. The measurement of the auxiliary qubit is
equivalent to a measurement of the DFS qubit.

\subsubsection{Building blocks of the quantum repeater using DFS qubits}\label{sec:IIIB2rep}

\begin{figure}[t]
\centering\includegraphics[]{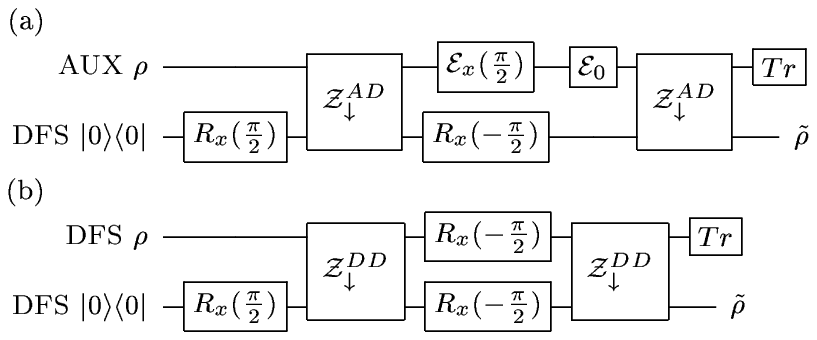} \vspace*{13pt}
\fcaption{%
  Quantum circuits for the transfer of (a) the state of an atomic
  qubit to a DFS qubit and (b) the sate of a DFS qubit to a DFS qubit.
  The operation $\mathcal{E}_0$ shown in (a) has no effect on the
  outcome of the state transfer, see text. In the noiseless case the
  output state $\tilde\rho$ would be equal to the input state
  $\rho$.}
\label{fig:transferrep}
\end{figure}

\begin{figure}[t]
\centering\includegraphics[]{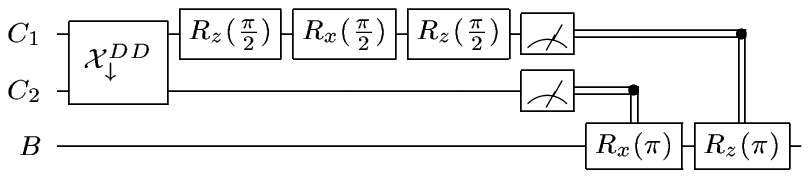} \vspace*{13pt}
\fcaption{%
  Quantum circuit for entanglement swapping involving only DFS qubits.
  The shown circuit corresponds to the standard protocol of
  teleporting the state of qubit $C_1$ to qubit $B$.}
\label{fig:DFSswappingrep}
\end{figure}

\begin{figure}[t]
\centering\includegraphics[]{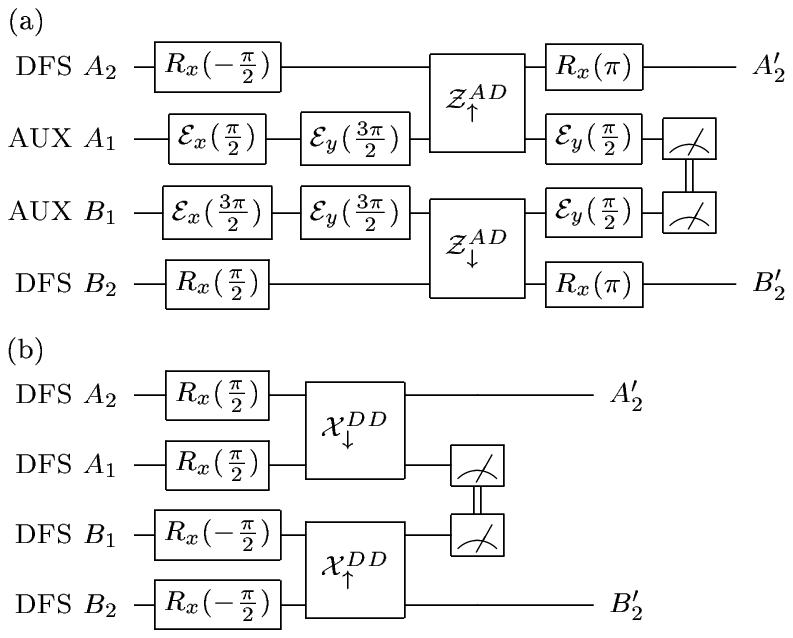} \vspace*{13pt}
\fcaption{%
  Quantum circuit for an entanglement purification step between (a)
  pairs of auxiliary and DFS qubits and (b) two pairs of DFS qubits.}
\label{fig:purifrep}
\end{figure}

The basic modules necessary to implement a quantum repeater
involving DFS qubits are shown in
Figs.~\ref{fig:transferrep}-\ref{fig:purifrep}. In the case of state
transfer shown in Fig.~\ref{fig:transferrep} we need two procedures,
namely a transfer from an auxiliary qubit to a DFS qubit and a
transfer of the state from one DFS qubit to another one.
In contrast to the state transfer between two DFS qubits (and also
between two unprotected qubits, see Fig.~\ref{fig:transfer_arep}),
which can theoretically be avoided, the transfer between auxiliary
(atomic) qubit and DFS qubit is necessary on the lowest level of the
quantum repeater. This is required since we assume that the state of
the flying qubit (typically a photon) is first transferred to an
atom, see Sec.~\ref{Sec:IIrep} and Ref.~\cite{Klein06}, and not
directly to a DFS qubit. The circuit shown in
Fig.~\ref{fig:transferrep}a includes a dephasing operation, which
accounts for the fact that the first qubit has to wait until the
$R_x(-\pi/2)$ gate on the second qubit is finished. This operation
is relatively slow ($\sim2.5\,$ms)~\cite{Klein06} and is thus much
slower than typical single qubit gates on atomic (auxiliary) qubits.
However, since the $\mathcal{E}_0$ operation commutes with the
$\mathcal{Z}^{AD}_\downarrow$ operation, this noise has no effect on
the outcome of the state transfer.

The quantum circuit for entanglement swapping with DFS qubits is
shown in Fig.~\ref{fig:DFSswappingrep}. The principle is the same as
described in Sec.~\ref{Sec:IIIA4rep}: The state of qubit $C_1$ is
teleported to qubit $B$ using the entanglement of the pair $C_2-B$.
Since quantum information can be stored in the DFS without losses,
waiting times during classical communications do not have any effect
if they do not exceed the coherence time of the DFS qubit (see
above). Thus qubit $A$ is omitted in this circuit since it would be
simply represented by a straight line.

Figure~\ref{fig:purifrep} shows the quantum circuit for entanglement
purification of (a) a pair of auxiliary qubits $A_1-B_1$ and a pair
of DFS qubits $A_2-B_2$, and (b) two pairs of DFS qubits $A_1-B_1$
and $A_2-B_2$. The major difference to the corresponding case of
unprotected qubits (Fig.~\ref{fig:purif_arep}) is again that the
waiting times during classical communication do not have any effect
if they are shorter than the coherence time of the DFS qubit.

\section{Results of simulations}\label{Sec:IVrep}

In this section, we present results of simulations of the full
nested purification protocol. For a better understanding we first
concentrate on its main components, namely entanglement swapping and
entanglement purification, as well as state transfer. As indicated
in Sec.~\ref{Sec:IIrep} the state transfer is theoretically not
necessary, however it turns out that its inclusion into the repeater
protocol does not reduce the final entanglement fidelity
significantly. Therefore, we include the state transfer into the
repeater protocol since in an experimental implementation it might
be easier to do so.
\begin{figure}[t]
\centering\includegraphics[]{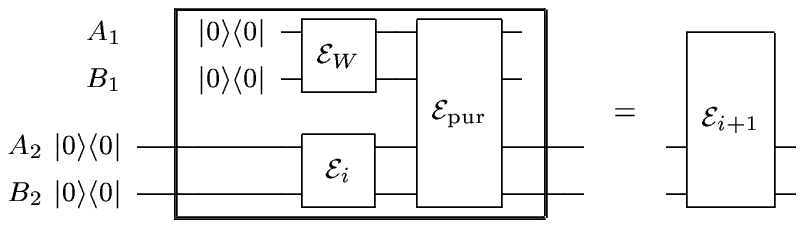} \vspace*{13pt}
\fcaption{%
  Illustration of the numerical method taking entanglement
  purification with Werner states as an example. The operation in the big box can be
  combined via process tomography to a single operation
  $\mathcal{E}_{i+1}$ given by a set of Kraus operators. Initially we
  have $\mathcal{E}_0 = \mathcal{E}_W$. The method is repeated $n$
  times corresponding to $n$ entanglement pumping processes. For more details
  see text.}
\label{fig:numeric_methodrep}
\end{figure}

We assume that the initial states are either Werner states
\begin{equation}
\rho_W = f_0\ket{\Phi_+}\bra{\Phi_+} + \frac{1-f_0}{3}(
\ket{\Phi_-}\bra{\Phi_-} + \ket{\Psi_+}\bra{\Psi_+} +
\ket{\Psi_-}\bra{\Psi_-} )
\end{equation}
or binary (mixture) states
\begin{equation}
\rho_B = f_0\ket{\Phi_+}\bra{\Phi_+} +
(1-f_0)\ket{\Phi_-}\bra{\Phi_-} \,,
\end{equation}
where
\begin{equation}
\ket{\Phi_{\pm}} = \frac{1}{\sqrt 2}( \ket{00}\pm\ket{11}  ) \,,
\qquad \ket{\Psi_{\pm}} = \frac{1}{\sqrt 2}( \ket{01}\pm\ket{10} )
\,.
\end{equation}
The entanglement fidelity of a state $\rho$ is defined as
\begin{equation}
f = \bra{\Phi_+}\rho\ket{\Phi_+} \,.
\end{equation}
In order to simulate the repeater and its constituents we developed
a program with a modular structure, i.e., it allows the simulation
of a quantum circuit by successively applying subroutines with mixed
states as input. Each of these subroutines corresponds to a quantum
operation (gates and measurements), which is represented by a Kraus
decomposition. Furthermore, we extensively use the fact that a
quantum network can be combined into one effective quantum
operation, the Kraus operators of which can be calculated by using
the quantum process tomography algorithm~\cite{Nielsen}.

We illustrate this method in more detail for the example of
entanglement purification with Werner states as shown in
Fig.~\ref{fig:numeric_methodrep}. The aim is to purify a qubit pair
$A_2-B_2$ via entanglement pumping using Werner states. The operation
$\mathcal{E}_W$ creates a Werner state out of the input state
$\ket{00}\bra{00}$ and the operation $\mathcal{E}_{pur}$ corresponds
to the actual entanglement purification circuit, for example as shown
in Fig.~\ref{fig:purif_arep}. The operation $\mathcal{E}_i$ is set
initially to $\mathcal{E}_0 = \mathcal{E}_W$. The circuit shown in
Fig.~\ref{fig:numeric_methodrep} then leads to an output state which
corresponds to the state after one entanglement pumping process. All
the operations inside the large box in
Fig.~\ref{fig:numeric_methodrep} can be combined to one effective
quantum operation $\mathcal{E}_1$, which acts on two qubits and which
can be determined via process to\-mo\-gra\-phy. The operation
$\mathcal{E}_1$ is then used instead of $\mathcal{E}_0$ in a
repetition of these steps and so forth. After $n$ iterations we get an
effective operation $\mathcal{E}_n$ which, when applied to the input
state $\ket{00}\bra{00}$, creates a state that we would get after $n$
entanglement pumping steps. By replacing the quantum operations inside
the large box in Fig.~\ref{fig:numeric_methodrep} appropriately, we
applied this method also to state transfer and entanglement swapping,
since in all cases all but two qubits are measured at the end of the
operation, i.e., also transfer and swapping can be represented by an
effective operation with two input and two output qubits. Ultimately,
the combination of state transfer, entanglement swapping, and
purification makes it possible to calculate an effective quantum
operation for the whole repeater, which transforms a given input state
(e.g.  $\ket{00}\bra{00}$) into the final state of the repeater.
\begin{figure}[t!]
\centering\includegraphics[]{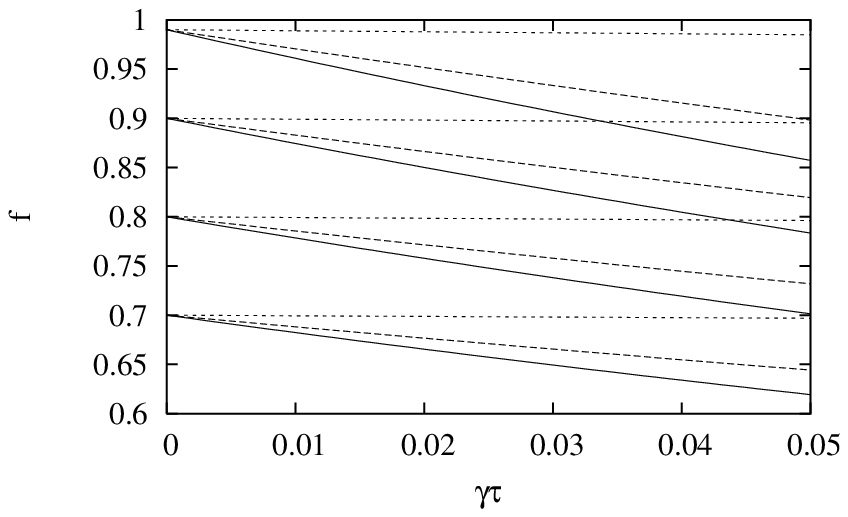} \vspace*{13pt}
\fcaption{%
  Fidelities after transfer of the state of a qubit pair depending on
  the noise parameter $\gamma$ ($\tau=1$ms). The initial states are
  Werner states with fidelities $f_0 = 0.7,0.8,0.9,0.99$ (bottom to
  top). Dotted lines correspond to the transfer of auxiliary
  (unprotected) qubits (cf. Fig.~\ref{fig:transfer_arep}), dashed lines
  correspond to a transfer from auxiliary to DFS qubits (cf.
  Fig.~\ref{fig:transferrep}a) and solid lines correspond to the transfer
  of DFS qubits (cf. Fig.~\ref{fig:transferrep}b).}
\label{fig:transfer_plotrep}
\end{figure}

For the examples shown in this section we set $\Omega = \Omega_x =
\Omega_y = \Omega_z = 2\pi\times 50$kHz, which means we assume that
a $2\pi$-rotation around the $x,y,z$ axes can be done in $10\mu$s.
The two particle interaction strength of Eq.~(\ref{eq:isingrep}) is
set to $\Omega_{zz}= 0.1\Omega$, i.e., the controlled-$Z$ gate
described in Fig.~\ref{fig:CZrep} takes $32.5\mu$s. Furthermore, we
assume an operation time for the controlled-$(-Z)$ operation between
an auxiliary qubit and a DFS qubit (see Fig.~\ref{fig:CZAUXrep}) of
$\tau = 1$ms, which corresponds to the gate operation times
calculated in Ref.~\cite{Klein06}. The measurement time is set to
$t_\mathrm{me}=10\mu$s and the measurement error is assumed to be
$1-\eta = 0.01$ unless otherwise stated.

Figure~\ref{fig:transfer_plotrep} shows the fidelity of the state of
an entangled particle pair after it was transfered to another qubit
pair versus $\gamma$ by means of the circuits described in the
previous section. The initial states are Werner states. The
corresponding plot for binary states (not shown) is very similar and
deviates from Fig.~\ref{fig:transfer_plotrep} appreciably only for
small fidelities. Assuming a coherence time of $1/\gamma = 100$ms,
the initial fidelity is reduced by $\sim 0.1\%$ in the case of a
transfer between auxiliary qubits, by $\sim 1-2\%$ for a transfer
from auxiliary to DFS qubits and by $\sim 2-3\%$ in the case of a
transfer between DFS qubits.
\begin{figure}[t!]
\centering
\includegraphics[]{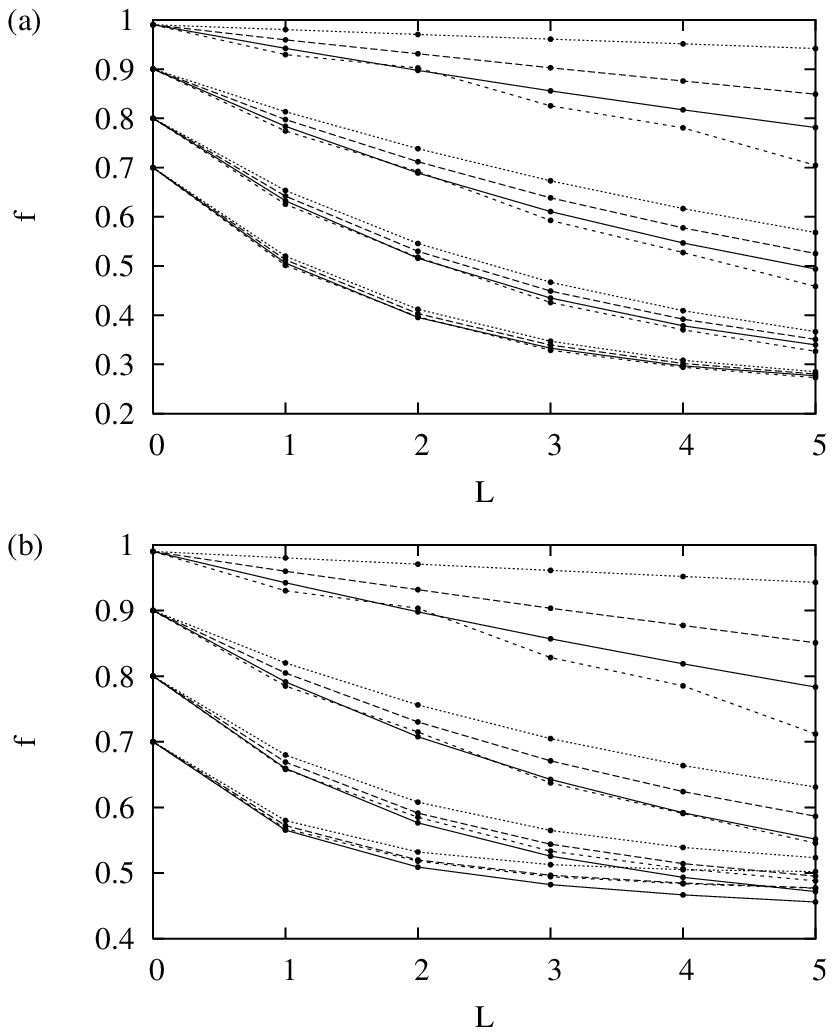}
\vspace*{13pt}
\fcaption{%
  Fidelities depending on the number of connection processes $L$ for
  $\gamma=1/100$ms. The qubit pairs are initially (a) in Werner
states and (b) in binary states with fidelities
$f_0=0.7,0.8,0.9,0.99$ (bottom to top). The solid lines correspond
to DFS qubits, and the short dashed and long dashed lines correspond
to auxiliary (unprotected) qubit pairs with distances of $l_0=1000$km
and $l_0=10$km, respectively. The dotted line corresponds to the
noiseless case, i.e., $\gamma = 0$, $\eta = 1$.}
\label{fig:swappingrep}
\end{figure}

The fidelities after connecting $L+1$ entangled qubit pairs (i.e.,
after $L$ connection processes) via entanglement swapping are shown in
Fig.~\ref{fig:swappingrep}. In these examples we assumed that there
are initially $L+1$ qubit pairs of fidelity $f_0$ and distance $l_0$
in a Werner state (Fig.~\ref{fig:swappingrep}a) or in a binary state
(Fig.~\ref{fig:swappingrep}b), which are connected according to the
method described in Sec.~\ref{Sec:IIIA4rep}. This implies that the
$i$th entangled pair with $i=1,2,\ldots,L+1$ has to wait an additional
time $\max\{0,|\lceil (L+1)/2\rceil-i|-1\}(l_0/c+t_\mathrm{sw})$,
until the connection process starts for this pair.  The final pair has
then a distance of $(L+1)l_0$.

As can be seen from these plots, entanglement swapping with
partially entangled states leads to a significant loss of fidelity,
in fact it has been shown that the fidelity decreases exponentially
in the noiseless case~\cite{Hartmann07}. The effect of noisy gate
operations and waiting times becomes less important for small
fidelities. For example, starting with two Werner pairs of fidelity
$80\%$ we lose about $15\%$ fidelity by connecting them. For short
distances a swapping procedure using unprotected qubits performs
generally better than one with DFS qubits due to the long gate
operation times in the decoherence-free case. However, for distances
$l_0$ larger than about $1000$km entanglement swapping with DFS
qubits becomes advantageous.

\begin{figure}[t!]
\centering\includegraphics[]{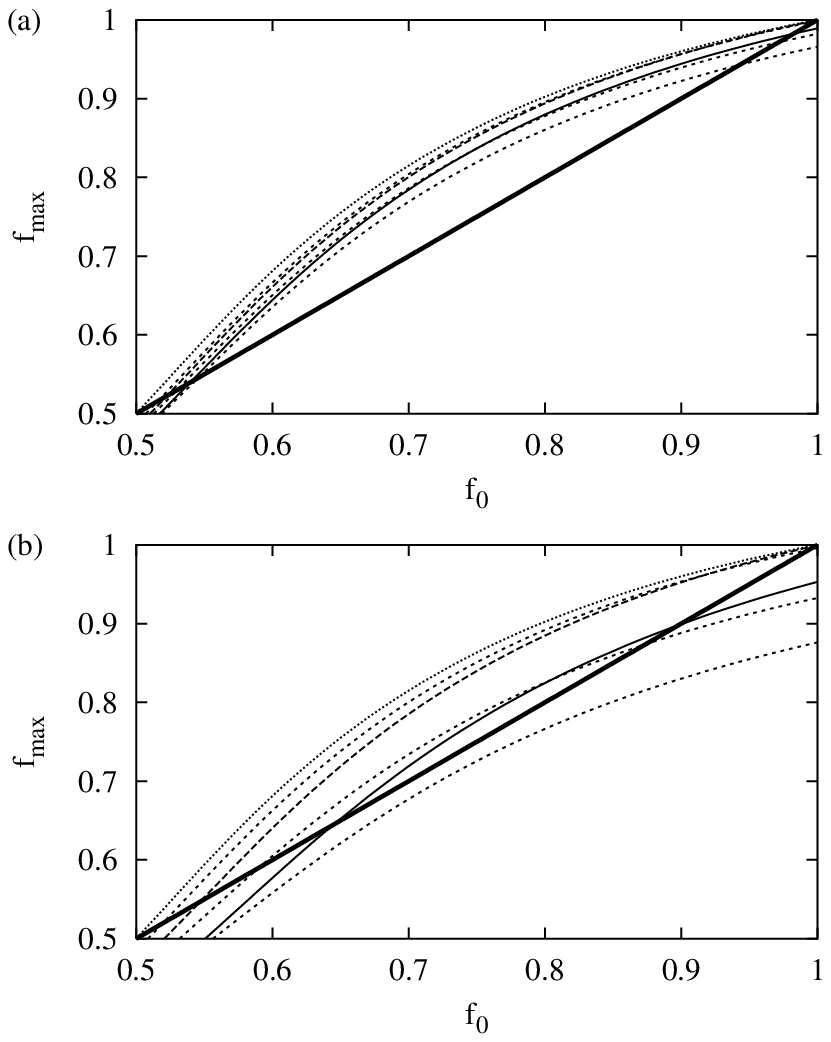} \vspace*{13pt}
\fcaption{%
  Maximal fidelity obtainable via entanglement purification using Werner states versus
  initial fidelity $f_0$, which is also the fidelity of the
  successively generated entangled pairs, for (a) $\gamma=1/100$ms and
  (b) $\gamma=1/25$ms. The bold, diagonal lines aid to read off
  whether
  entanglement is gained or lost, see text. The thin, solid lines
  correspond to purification using DFS qubits, and the short dashed
  lines correspond to purification using unprotected qubits for
  $l_0=10$km, 500km, 1000km (top to bottom). The long dashed line was
  obtained by purifying a DFS qubit pair with an auxiliary
  (unprotected) qubit pair. As a reference we also plotted the
  corresponding result for the noiseless case (dotted lines). }
\label{fig:purif_wernrep}
\end{figure}
\begin{figure}[t]
\centering\includegraphics[]{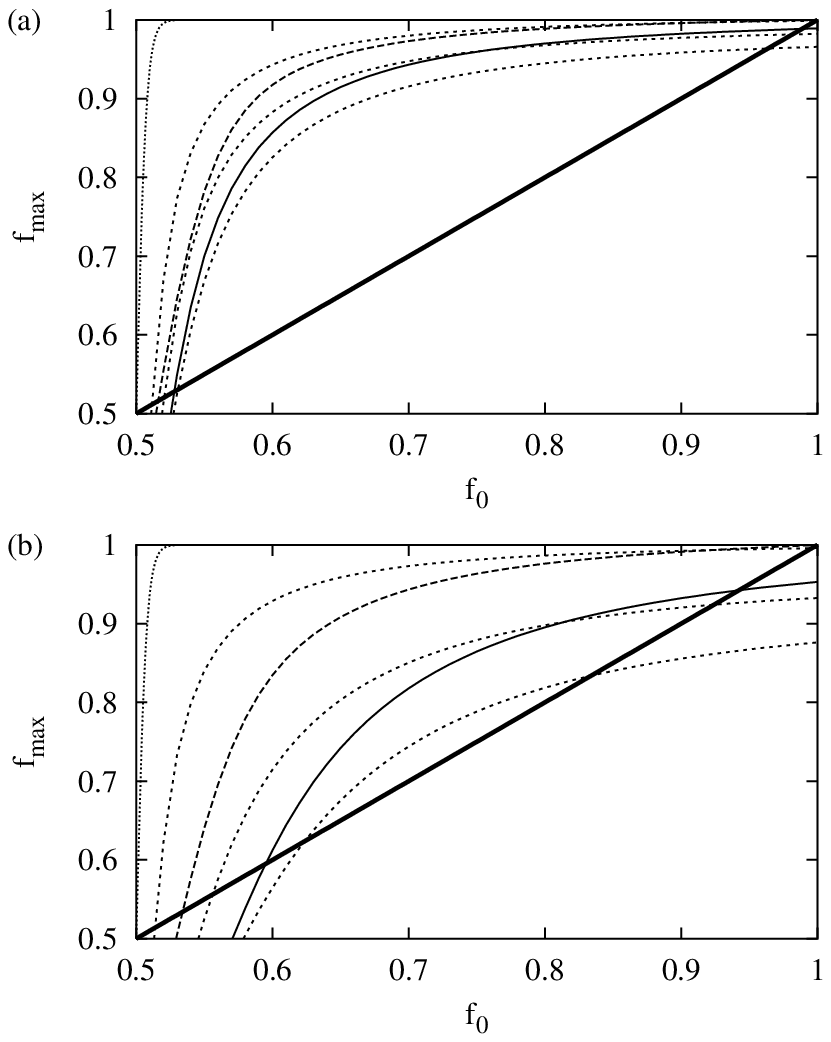} \vspace*{13pt}
\fcaption{%
Same as in Fig.~\ref{fig:purif_wernrep} but using binary states
instead of Werner states.} \label{fig:purif_binrep}
\end{figure}

The effect of entanglement purification (using entanglement pumping)
of an entangled qubit pair, which has initially the fidelity $f_0$,
is shown in Fig.~\ref{fig:purif_wernrep} and
Fig.~\ref{fig:purif_binrep}. In particular, we calculated the
maximally reachable fidelity $f_{\mathrm{max}}$ after a large number
of successful purification steps depending on $f_0$, which is also
the fidelity of the successively provided additional pairs used for
entanglement pumping. In Figs.~\ref{fig:purif_wernrep}
and~\ref{fig:purif_binrep} the initial pair and the additional pairs
are Werner states and binary states, respectively. Whenever the
curves are above the bold diagonal line we gain fidelity, otherwise
the fidelity is decreased during the process. These figures
illustrate again that the DFS scheme becomes better than the scheme
based on unprotected qubits at a distance of $l_0 \approx 500$km.
For $\gamma=1/25$ms and a distance of $l_0 = 1000$km (lowest short
dashed line in Fig.~\ref{fig:purif_wernrep}b) purification would not
be possible at all for unprotected qubits. Moreover, binary states
perform generally better than Werner states~\cite{Duer99}. Also
shown in these figures are the maximal fidelities for purification
between an auxiliary qubit and a DFS qubit (long dashed lines),
which is needed on the first level of the DFS repeater, compare
Fig.~\ref{fig:purifrep}a. Note that in this case the maximally
reachable fidelity, i.e., the point where the curves intersect the
diagonal line in the upper right corner of the plots, is very close
to one even for small coherence times $1/\gamma$. The points where
the curves intersect the bold diagonal line in the lower left corner
correspond to the purification threshold below which no purification
is possible.

\begin{figure}[t!]
\centering\includegraphics[]{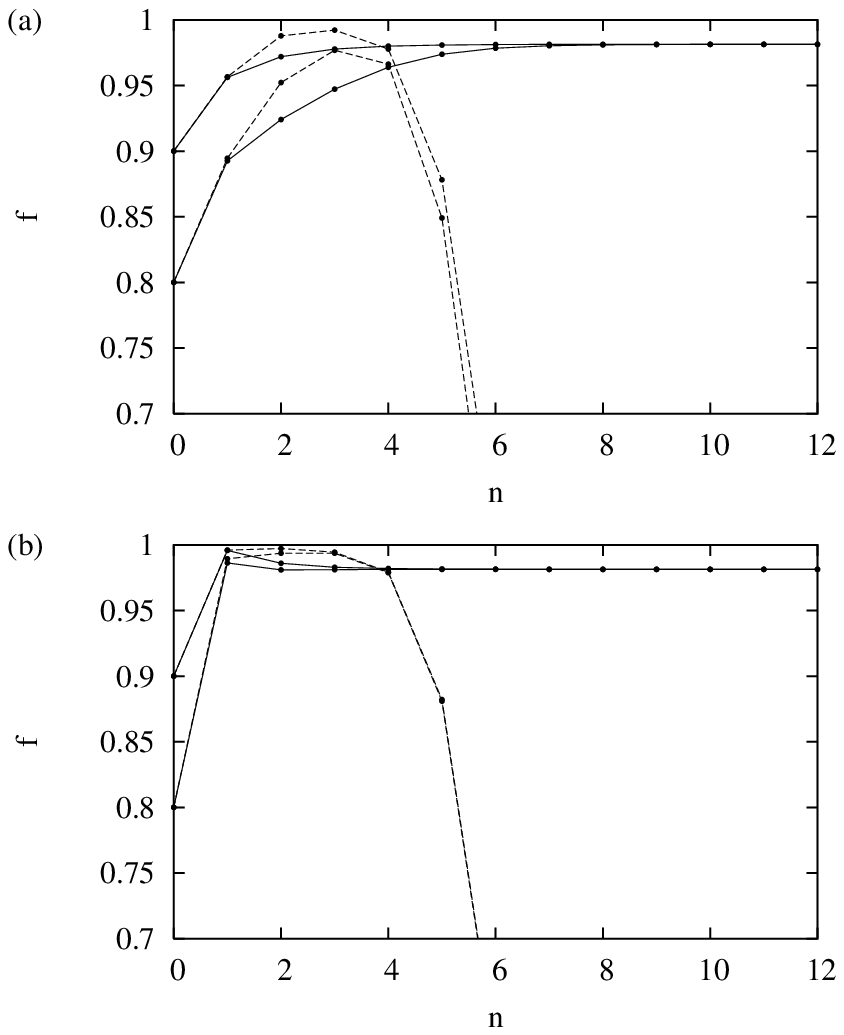} \vspace*{13pt}
\fcaption{%
  Fidelity $f$ of an entangled pair created by a quantum repeater
  using (a) Werner states and (b) binary states on the lowest level
  depending on the repeater level $n$. The solid lines correspond to a
  DFS repeater and the dashed lines to a repeater based on unprotected
  qubits. The fidelities of the initially created pairs are
  $f_0=0.8,0.9$ (bottom to top), $l_0=10$km and $\gamma=1/100$ms. For further
  parameters see text.}
\label{fig:repeaterrep}
\end{figure}

As described in Sec.~\ref{Sec:IIrep}, the quantum repeater protocol
we use in this paper is a nested arrangement of entanglement
purification and entanglement swapping. The DFS repeater suffers
from long gate operation times, which particularly affects the noisy
two qubit gate. Quantum gates based on unprotected qubits are
significantly faster, however, due to long waiting times during the
repeater protocol in the case of long distances, the involved memory
qubits are strongly prone to decoherence. In
Fig.~\ref{fig:repeaterrep} we show an example which compares these
two cases.
We calculated the fidelity of the entangled pair generated by a
quantum repeater with $n$ levels. On each level we perform 5
purification steps and perform one connection (i.e., $L_j=1$) except
for the first level where no connection is done (i.e., $L_1=0$). For
the repeater based on unprotected qubits we assume $t_0=10\mu$s. The
distance of the generated entangled pair on level $n$ scales like
$S_n=2^{n-1}l_0$. If we take for example $l_0 = 10$km and $n=12$, we
get a distance of 20480km. Clearly, we do not suggest that the
distance over which entangled pairs can be distributed with the DFS
quantum repeater is unlimited.  The DFS we use protects only against
noise given by Eq.~(\ref{decoherence_fieldrep}) and is for example
not immune to fluctuating inhomogeneous fields. Therefore, the
``decoherence free'' subspace ultimately has a finite coherence time
which, however, can be very long (see Sec.~\ref{SecIIIB1rep}). In
particular, we expect it to exceed the time necessary to generate an
entangled pair on an intercontinental distance, which is on the
order of tens of seconds~\cite{Duer99}. As can be seen from
Fig.~\ref{fig:repeaterrep}, the DFS repeater outperforms the
repeater based on unprotected qubits already on repeater level 4
which corresponds to 80km in the above example. The final fidelity
of the entangled pairs produced by the DFS repeater is $f=98.1\%$.
The waiting times, which are relevant for the repeater based on
unprotected qubits, are calculated according to
Eqs.~(\ref{Eq:Waitpurrep}) and~(\ref{Eq:Waittrrep}), i.e., they
represent a lower bound. We emphasise here that the strategy (i.e.,
the choice of the $L_j$, the number of purification steps on a
repeater level, the distance $l_0$ of the initially created pairs
etc.) used in this example might not be the most optimal one. A
systematic approach to this problem, albeit with a different error
model, can be found in \cite{Hartmann07}. However, it was found in
this reference that even with an optimised strategy and reasonable
errors intercontinental distances can not be reached by a repeater
using entanglement pumping. Therefore, a number of alterations to
the repeater protocol have been proposed to increase the
distance~\cite{Hartmann07}. Our results show that improving the
quantum memory by employing a decoherence free subspace (and not
changing the repeater protocol) gives an alternative method to reach
intercontinental distances.

\section{Conclusions\label{Sec:Conclrep}}
 In the present paper we described in detail the implementation of
a quantum repeater based on DFS quantum memories, i.e., the qubits
at the repeater nodes are represented by two states of a DFS
consisting of four physical qubits, which can be manipulated as
proposed in Ref.~\cite{Klein06}. We showed that the distribution of
entangled pairs over long distances is possible with our setup. We
simulated the DFS repeater as well as a repeater based on
unprotected qubits using realistic parameters, and demonstrated that
the DFS scheme outperforms the scheme based on unprotected qubits if
waiting times due to classical communication become too large. The
implementation of a repeater based on a reliable memory, even in the
case of slow and faulty gate operations as in the example we
considered, would thus offer the possibility for long distance
quantum communication.

In future work one might enhance the performance of the quantum
repeater even further by conceiving hybrid architectures, which
combine the advantages of fast schemes based on unprotected qubits and
schemes involving DFS qubits. For example, one could use the DFS
qubits merely as a memory and quantum information is processed with
unprotected qubits. The memory could be accessed via state transfer
mechanisms as it is discussed in the present paper acting as an
interface between memory and processing qubits. Fast gate operations
would then be performed on and between unprotected qubits. A further
option along the lines of these ideas is based on the observation that
the maximally reachable fidelity $f_{\mathrm{max}}$ of purification of
a DFS qubit pair with unprotected qubit pairs is larger than
purification using only DFS qubit pairs (see long dashed lines in
Figs.~\ref{fig:purif_wernrep} and~\ref{fig:purif_binrep}): On the
final repeater level(s) we could therefore transfer the state of the
DFS qubit pair to an unprotected qubit pair and use this in turn to
purify another DFS qubit pair. If the loss in fidelity induced by the
state transfer is not too high the results presented in
Figs.~\ref{fig:purif_wernrep} and~\ref{fig:purif_binrep} suggest that
this could lead to final fidelities exceeding those obtained from
using exclusively DFS qubits on higher repeater levels.

\nonumsection{Acknowledgements}

This research was supported by a Marie Curie Intra-European
Fellowship within the 6th European Community Framework Programme
(`RAQUIN'). This work was also supported by the EPSRC (UK) through
the QIP IRC (GR/S82176/01) and project EP/C51933/1 and by the EU
through the STREP project OLAQUI. A.K.~acknowledges financial
support from the Keble Association.

\end{document}